\documentstyle[titlepage,twoside,11pt]{article}
\begin{document}
\pagestyle{myheadings}
\markboth{\it A. F. Nicholson \& D. C. Kennedy}{\it A. F. Nicholson \& 
D. C. Kennedy}
\def\call{{\cal L}}

\def\be{\begin{equation}}
\def\bq{\begin{eqnarray}}
\def\ee{\end{equation}}
\def\eq{\end{eqnarray}}

\def\GSW{{\rm GSW}}
\def\QED{{\rm QED}}
\def\QCD{{\rm QCD}}
\def\rsl{{\rm sl}}
\def\phys{{\rm phys}}
\def\noas{\noalign{\smallskip}}
\def\tes{\textstyle}
\def\ite{{\sl et al.\/}}
\def\suu{$SU(2)_L\times U(1)$}
\def\call{{\cal L}}
\def\calm{{\cal M}}
\def\calS{{\cal S}}
\def\cosec{\,{\rm cosec}\,}
\def\eps{\epsilon}
\def\th{\theta}
\def\mn{{\mu\nu}}
\def\cald{{\cal D}}
\def\calh{{\cal H}}
\def\thf{\textstyle{1\over 2}}
\def\tqr{\textstyle{1\over 4}}
\def\tos{\textstyle{1\over 6}}
\def\ttt{\textstyle{2\over 3}}
\def\tot{\textstyle{1\over 3}}
\def\bare{\scriptstyle{{\rm bare}}}
\def\c@ncel#1#2{\ooalign{$\hfil#1\mkern1mu/\hfil$\crcr$#1#2$}}
\def\slp{\partial\!\!\!/}
\def\slnp{p\!\!\!/ }
\def\slA{A\!\!\!\!/ \,}  
\def\slB{B\!\!\!\!/ \,}  
\def\slW{W\!\!\!\!\!/ \,}

\begin{titlepage}
\begin{flushright}UF-IFT-HEP-96-29\\ Original: February 1997\\
Revised: June 1999\end{flushright}
\ \\
\begin{center}
{\bf GAUGE BOSON AND FERMION MASSES}\\
{\bf WITHOUT A HIGGS FIELD}\\
\ \\ \ \\
Angus F.~Nicholson\footnote{Deceased}\\
{\it Department of Theoretical Physics, IAS}\\
{\it The Australian National University, Canberra, ACT 0200, 
Australia}\\
\ \\ \ \\
Dallas C.~Kennedy\footnote{e-mail: kennedy@phys.ufl.edu}\\
{\it Department of Physics, University of Florida,}\\
{\it Gainesville, Florida 32611, USA}\\
\ \\ \ \\
\ \\
{\bf ABSTRACT}

\end{center}

A simple, anomaly-free chiral gauge theory can be perturbatively 
quantised and renormalised in such a way as to generate fermion and 
gauge boson masses.  This development exploits certain freedoms 
inherent in choosing the unperturbed Lagrangian and in the 
renormalisation procedure. Apart from its intrinsic interest, 
such a mechanism might be employed in electroweak 
gauge theory to generate fermion and gauge boson masses without a 
Higgs sector.
\ \\ 
\begin{center}
Submitted to {\bf International Journal of Modern Physics A}
\end{center}

\end{titlepage}

\addtocounter{page}{1}

\section{Introduction}

There is a widespread, plausible view that the Standard Model is 
a low-energy effective theory and that the Higgs boson might not 
exist [1].
A Higgs sector introduces naturalness, triviality and cosmological 
problems [2].  The simple model 
given in this paper illustrates a perturbative method 
for giving mass to gauge bosons and fermions
with chiral couplings which does not require 
Higgs fields or new fermions.

The mass mechanism presented here might be used in a 
fundamental theory or in a low-energy effective field theory. An 
effective theory need not be renormalisable [3].
Our model gives mass to the gauge boson on 
renormalisation to one loop, is renormalisable to several loops 
or orders, and appears to possess QED-like renormalisability to 
all orders. The model branches into three cases, in which the masses 
are $O(g^0)$, $O(g^1)$ (like $m_W$, $m_Z$ in standard electroweak 
theory) and $O(g^2)$.  The mechanism seems to be applicable to any
chiral gauge theory.

The effective Lagrangian density of the model, including gauge-fixing 
and ghost terms, is 
\begin{eqnarray} \call_{\rm eff} &= -
{\tqr} Z_3(\partial_\mu A_\nu -\partial_\nu A_\mu)^2 - 
(2\xi)^{-1} Z_4(\partial_\mu A^\mu)^2 
+(\partial_\mu c^*)\partial^\mu c \nonumber \\ 
&\qquad+ 
Z_{2L}\hbar\bar\psi_L i\slp \psi_L +Z_{2R}\hbar \bar\psi_R i\slp 
\psi_R -\hat g\bar\psi \gamma^\mu [Z_1+fZ_{1(5)} \gamma^5]\psi 
A_\mu \nonumber \\ 
&\qquad+ Z_{6L}\hbar \bar\eta_L i\slp \eta_L 
+Z_{6R}\hbar \bar\eta_R i\slp \eta_R +\hat g \bar\eta \gamma^\mu 
[Z_5 +fZ_{5(5)}\gamma^5] \eta A_\mu, \end{eqnarray} which has 
been obtained from a chiral- and gauge-invariant classical 
$\call$ in the path integral formalism.  A renormalisation 
$Z_4$ of the gauge-fixing term is allowed. We have included 
$\hbar$, dimensionless but not yet numerically 1, 
since $\hbar^{-1}$ is the expansion parameter in our Case~B (see 
section~2).\par 

There are no mass terms in $A_\mu A^\mu$, $\bar\psi\psi$ or 
$\bar\eta\eta$ in $\call_{\rm eff}$, and in section~2 we see that 
the effective action $S_{\rm eff}=\int d^4 x \call_{\rm eff}$ is 
invariant under $U(1)_{L,R}$ BRS transformations.
For $f\ne 0$, its $U(1)_{L,R}$ invariance does not 
permit $\call$ or $\call_{ \rm eff}$ any mass terms. It might 
then appear that $\call_{\rm eff}$ must define a massless theory.
However, we recall nonperturbative 
counterexamples. The standard electroweak Lagrangian in its 
initial form does not contain vector boson mass terms of the form 
$M^2 A_\mu A^\mu$, yet the theory gives massive $W$, $Z$ bosons by
spontaneous symmetry breaking (SSB). 
For the QED Lagrangian \begin{eqnarray} \tilde\call = -{\tqr} 
(\partial_\mu A_\nu -\partial_\nu A_\mu)^2 
 +\bar\psi(i\slp -e\slA\,)\psi
\end{eqnarray}
in $1+1$ spacetime dimensions, the Schwinger model [4] has an exact 
solution in which the only physical particle is a massive vector 
boson [5,6], the mass arising from quantum corrections.
Other nonperturbative examples with initially massless fermions and
dynamical breaking of chiral symmetry are known.
Nambu and Jona-Lasinio (NJL) [7] 
described nucleons with a four-fermion interaction,
a mechanism natural with massive gauge boson exchange.
Baker {\sl et al.} and Green {\sl et al.} [8] obtained from massless QED 
asymptotic
and full forms, respectively, of a propagator approximating 
the standard form for a massive fermion.
Jackiw and Johnson [9] and Cornwall and Norton [10] 
generated fermion and gauge masses by massless bound 
$f\bar f$ excitations.  More recently,
Roberts and others [11,12,13] have obtained quark self-energies
from truncated Dyson--Schwinger equations for QCD, 
breaking the chiral symmetry dynamically. Gusynin, Miransky and 
others [14,15] have applied nonperturbative methods to the NJL model 
and QED and obtained 
dynamical fermion masses for an arbitrarily weak attraction
between fermions.
In contrast to these nonperturbative developments, we show that 
$\call_{\rm eff}$ can give fermions and gauge bosons masses 
{\it perturbatively.} We make an 
unconventional ansatz for $\call_0$ in the decomposition 
$\call_{\rm eff}=\call_0 +\call_1$ and renormalise in a novel way.

Gauge symmetry is replaced by BRS symmetry because 
$\call_{\rm eff}$ contains a gauge-fixing term and ghosts. Then BRS 
symmetry is broken upon making the decomposition 
$\call_{\rm eff} = \call_0 +\call_1$. 
While the effective action $S_{\rm eff} = \int d^4 x \call_{\rm 
eff}$ is BRS-invariant in QED, QCD and this model, the partial-actions 
$S_{0,1} = \int d^4 x \call_{0,1}$ are not. 
This lack of gauge symmetry in $\call_0$ does not 
matter, provided that the choice of $\call_0$ gives 
$S$-matrix elements independent of the 
gauge (an issue not fully resolved [16]). In this model $\call_0$ 
also does not possess the chiral 
invariance of $\call_{\rm eff}$. However, the propagators and 
vertices that we obtain, and so the $S$-matrix elements, do not 
depend on the chirality phase, in the way that 
propagators in QED and QCD usually depend on a gauge parameter 
$\xi$.  

It is conventional to take as $\call_0$ the quadratic 
part $\call_{\rm quad}$ of a given $\call$, 
rather than to quantise from
another $\call_0$, i.e., from $\call_0 =\call_{\rm quad}+\chi .$ 
For example, given the form (2), it would 
be normal to take $\tilde\call_0$ to be the 
quadratic part of $\tilde\call$ and expect the result to be 
massless QED. Yet the examples cited above show that a theory
massless in its quadratic part can nonetheless generate masses
dynamically once interactions are introduced.
We take the view that any choice of $\call_0$ that 
leads to a self-consistent model is legitimate, especially if the 
mass mechanism proposed gives promise of being applicable in a 
physical theory, such as an electroweak theory without a Higgs 
sector.

We use a decomposition of $\call$ in which $\call_0$ contains 
fermion and gauge boson mass terms.  The mass term 
\begin{eqnarray} \chi = {\thf} \hbar^{-2} m_A^2 A_\mu A^\mu -
m_1\bar\psi\psi-m_2\bar\eta\eta, \end{eqnarray} is placed in $\call_0$ and 
$-\chi$ in $\call_1$, so that $\call_{\rm eff}$ is unchanged and 
the $U(1)_{L,R}^{}$ invariance is unaffected. 
Splitting a zero term in $\call_{\rm eff}$ to obtain 
$\chi,-\chi$ in $\call_0,\call_1$ is akin to 
splitting $m_0\bar\psi\psi$ into $m_{\rm phys} \bar\psi\psi$ 
and $\delta m \bar\psi\psi$ in QED.
We make the choice 
\begin{eqnarray}\call_0 &= -{\tqr} (\partial_\mu A_\nu -
\partial_\nu A_\mu)^2 -(2\xi)^{-1} (\partial_\mu A^\mu)^2 
+{\thf}\hbar^{-2} m_A^2 A_\mu A^\mu +(\partial_\mu c^*) \partial^\mu 
c \nonumber \\ 
&\qquad+ \bar\psi(i\hbar\slp-m_1)\psi 
+\bar\eta(i\hbar\slp -m_2)\eta \end{eqnarray} 
(since $\psi_L 
i\slp \psi_L +\bar\psi_R i\slp \psi_R =\bar\psi i\slp\psi$, and 
similarly for $\eta$), 
which generates massive particles, and show that we can regenerate 
dynamically the masses $m_1$, $m_2$, and 
$m_A$ or generate a complex mass for the boson (when it is 
unstable), to obtain a self-consistent theory. 
We begin from $\call_{\rm eff}$, which by itself does not necessarily
exhibit the physical spectrum.

It might appear that the $(\chi,-\chi)$ step must be nugatory, since 
the boson and fermion 
mass terms from $\chi$ in $\call_0$ and $-\chi$ in $\call_1$ 
cancel in the denominators of the full, improper propagators, 
e.g., in $\{\slnp -m -[-m+\Sigma(\slnp) +c.t.]\}.$ However, $m$ 
contributes to $\Sigma(\slnp)$, and this leads to nonzero 
renormalised fermion masses. The $m_1$, $m_2$ terms 
in the fermion propagators that follow from (4) (in section~2) 
are essential for the generation of a $g_{\mn}\rho(k^2)$ term in 
the boson self-energy tensor $\pi_{\mn}(k)$, which is 
\be
\pi_{\mn}(k)  = (k_\mu k_\nu -g_{\mn}k^2) \pi(k^2)+k_\mu k_\nu 
\tau(k^2) +g_{\mn} \rho (k^2), 
\ee
where, to avoid 
ambiguity, $\rho(k^2)$ is defined not to contain a factor $k^2$.
It is this term that leads to the mass of the gauge boson in our model. 
Similar mass terms appear in $\pi_{\mn W}$, $\pi_{\mn Z}$ in 
standard electroweak theory [17].  The nonzero 
fermion and boson mass ansatz can be maintained self-consistently: 
the loops 
regenerate the masses, or generate a complex boson mass, so that they 
are quantum-field-theoretical in origin. 

So that the series for the full boson and fermion propagators can
be summed in the usual way, correctly to any given power of the 
expansion parameter, each term in $\call_1$, and so
$m_1,\ m_2,\ m_A$ in $-\chi$, must be proportional to 
a positive power of that parameter. In Case A1 and Case A2 the 
expansion parameter is $g$. In Case A1 we take 
\begin{eqnarray} 
 {\rm A1:} \quad m_1 = \beta_1 g, \quad m_2 = \beta_2 g,\quad m_A 
=\beta_A g, \end{eqnarray}  
and on renormalisation obtain a boson mass of $O(g)$.  We recall 
that $m_W$, $m_Z$ are $O(g)$ in standard electroweak theory, and 
that the boson mass is $e\pi^{-1/2}$ in the Schwinger model.  In 
Case A2 we take 
\begin{eqnarray} 
{\rm A2:} \quad m_1 = \beta_1 g^2, \quad m_2 = \beta_2 g^2,\quad m_A 
=\beta_A g^2, \end{eqnarray}  
and obtain a boson mass of $O(g^2)$. In both these cases we could 
put $\hbar=1$ at the beginning. In Case B the expansion parameter 
is $w=\hbar^{-1}$ (see section 2) and we put 
$\hbar =1$ at the end of the renormalisation. We take 
\begin{eqnarray} 
{\rm B:} \quad m_1 = \beta_1 w, \quad m_2 = \beta_2 w,\quad m_A 
=\beta_A w, \end{eqnarray}  
and the final boson mass is $O(w) = O(w g^0)$, i.e., $O(1)$ 
when we put $\hbar =1$. Thus our three cases cover masses of $O(1)$, 
$O(g)$ and $O(g^2)$.  

We employ dimensional regularisation, with $\gamma^5$ 
totally anticommuting in $d=4-2\epsilon$ dimensions 
[18,19,20], a standard 
regularisation method [17,21,22,23,24].
At one loop, the divergent part of $\rho(k^2)$ is 
\begin{eqnarray} \rho_{2\eps} = {\hbar^{-2} g^2 f^2\over 2\pi^2\eps} 
(m_1^2 +m_2^2).  \end{eqnarray} Since no gauge boson mass 
counterterm can appear in $\call_{\rm eff}$, $\rho_{2\eps}$ cannot be 
cancelled. Further, $\rho_{2\eps}$ vanishes as $f\rightarrow$ 0: 
the generation of gauge boson mass is chiral 
in this model.

If the renormalisation gives the boson a mass $m_{AR}$ such that the 
boson is stable, $m_{AR}<2m_1$, then the particles are stable, 
since the interactions in $\call_1$ ensure that the 
fermions are stable. Then unitarity requires that the LSZ S-matrix
reduction formula [25] hold in the renormalised theory, so that the
renormalised masses equal the masses in $\call_0$, i.e., $m_{AR}=
m_A$, $m_{1R}=m_1$, $m_{2R}=m_2$.

However, the only known massive elementary bosons are the $W$ and the
$Z$, so it is the unstable boson case that is of 
physical interest. Then the boson mass $m_{AR}$ must be complex (e.g., 
for the $Z$ boson [26,27]).
But the usual perturbative or LSZ [28] formalisms do not 
accommodate unstable particles in a consistent way [29,30].  
Questions of gauge 
invariance, unitarity, renormalisation and the complex $Z$ mass 
pole are discussed in [31,32,33,34].
In this paper we use the standard perturbative 
formalism when there is an unstable particle present, but if our 
mass mechanism were used in a fundamental theory containing 
unstable particles, this formalism could not treat unitarity 
consistently.  Moreover, predictions from an effective Lagrangian must be
checked for unitarity at high energies [2]. We 
postpone further consideration of unitarity to a later publication.  

\section{$U(1)$ and Chiral Invariance, Propagators} 

Since we work in $d=4-2\eps$ dimensions, $\hat g$ in (1) is $\hat 
g = g\mu^\eps$, where $g$ is dimensionless and $\mu$ is a scale 
mass; however, hereafter we drop this distinction, and
$\hat g$ is to be understood where 
appropriate. In the factors $Z_i = 1-c_i$ in (1), the $c_i$ are 
counterterm parameters.  The primary interaction term is 
\begin{eqnarray}  \call_{1P} &=  -g \bar\psi \gamma^\mu (1+f 
\gamma^5) \psi A_\mu +g\bar\eta \gamma^\mu(1+f\gamma^5)\eta A_\mu 
\nonumber \\ &= -g [(1-f)\bar\psi_L \gamma^\mu \psi_L +(1+f) 
\bar\psi_R\gamma^\mu \psi_R] A_\mu \nonumber \\ &\qquad + g[(1-
f)\bar\eta_L\gamma_\mu \eta_L +(1+f) \bar\eta_R \gamma^\mu 
\eta_R] A_\mu.  \end{eqnarray} The two anomalous  
fermion triangle divergences cancel, because of the $-g,g$ 
couplings to $A_\mu$, and that the fermion quadrangle loops 
(applying an argument similar to that of [35]
for QED) and higher polygon loops are convergent. In 
the limit of $f\rightarrow$ 0, the theory becomes
two-fermion QED with zero gauge boson mass.

It is straightforward to transform in the usual way to $\xi^0 =  
Z_4^{-1} Z_3\xi$, \ $A^0_\mu = Z_3^{1/2} A_\mu$, $\psi^0_{L,R} = 
Z^{1/2}_{2L,R} \psi_{L,R}$, \ $\eta^0_{L,R} = Z^{1/2}_{6L,R} 
\eta_{L,R}$ and 
show that the action $S=\int d^4 x\call_{\rm eff}$ is invariant 
under the $U(1)_{L,R}$ BRS transformation 
\begin{eqnarray} \left. 
\begin{array}{ll} A_\mu^0 & \to A_\mu^0+ \tau \partial_\mu c \\ 
c^* & \to c^*-{\displaystyle{1\over \xi^0}} \tau \partial^\mu 
A_\mu^0, \quad 
c\to c \\ 
\psi_L^0 & \to (1-i \hbar^{-1} g^0_{1L}\tau c)\psi^0_L, \quad 
\eta^0_L \to (1+i \hbar^{-1} g^0_{5L} \tau c)\eta^0_L \\ 
\psi^0_R & \to  (1-i\hbar^{-1}  g^0_{1R}\tau c) \psi^0_R, \quad 
\eta^0_R \to (1+i \hbar^{-1} g^0_{5R} \tau 
c) \eta^0_R \end{array} \qquad\right\} \end{eqnarray} where 
$\tau$ is a Grassmann number and 
\begin{eqnarray} \left. 
\begin{array}{ll} g^0_{1L} & = (Z_1 -f Z_{1(5)}) Z_{2L}^{-1} 
Z_3^{-1/2} g\\ 
g^0_{1R} & = (Z_1 +f Z_{1(5)}) Z_{2R}^{-1} Z_3^{-
1/2} g\\ 
g^0_{5L} & = (Z_5 -f Z_{5(5)}) Z_{6L}^{-1} Z_3^{-1/2} 
g\\ 
g^0_{5R} & = (Z_5 +fZ_{5(5)}) Z_{6R}^{-1} Z_3^{-1/2} g. 
\end{array} 
\qquad\right\} \end{eqnarray} Accordingly, we have a 
$U(1)_{L,R}$ gauge theory. In addition, $\call$ is invariant 
under the chirality transformation $\psi \to e^{i\alpha\gamma_5} 
\psi ,$\ $\eta\to e^{i\beta\gamma_5}\eta$, with $\alpha$, $\beta$ 
constants. The $U(1)_{L,R}$ invariance, with $f\ne 0$, forbids mass terms.
Since $c^*,c$ do not couple to $A_\mu$, we omit them from this 
point onwards.  The BRS and chiral symmetries imply the usual Ward
identities for (1).

Placing the mass term $\chi$ given by 
(3) in $\call_0$ and $-\chi$ in $\call_1$ does not affect the 
invariances of $\call$ and the action. $\call_0$ does 
not possess the chiral invariance of $\call$, and, similarly 
to the situation in QED and QCD, the partial action $S_0 = \int d^4 
x\,\call_0$ does not possess the $U(1)_{L,R}$ invariance of $S$.  

If the renormalisation is performed so 
as to give a stable boson of mass $m_{AR}$, we must have $ 
m_{AR}=m_A$ and 
the renormalised fermion masses must be $m_{1R}=m_1$, 
$m_{2R}=m_2$. If the boson is to be unstable, 
the case of physical interest, then it must have a complex mass 
$ m_{AR}$, which cannot equal a real $m_A$.
Quantisation with any value for $m_A$ gives a 
set of basis states which include a non-degenerate vacuum 
that possesses the symmetries of $\call_0$.

In treatments of QED in which $\hbar$ and $c$ are dimensionful
[36], $\call_0$ for a fermion is $\hbar c\bar\psi i\slp\psi -m 
c^2 \bar\psi\psi$, and diagrams are generated by the action of $(\hbar 
c)^{-1}\call_1$. Taking $c=1$ and $\hbar$ dimensionless but not yet 1 
gives the forms of the fermionic parts in $\call_{\rm eff}$, (1), and 
$\call_0$, (4), and at each vertex we have $g\hbar^{-1}$. 
We obtain from (4), by the path integral method or 
canonical quantisation, the usual Fourier-transformed propagators
\begin{eqnarray} i D_{\mn}(k) 
 & = & {-i\hbar \over k^2-\mu^2+i\eps'} \bigg[ g_{\mn} +(\xi-1){k_\mu 
k_\nu\over k^2-\xi \mu^2+i\eps'} 
\bigg], \\ 
i S_{F\psi,\eta}(p) 
 & = & i [\slnp -\kappa_{1,2}+i\eps']^{-1},
\end{eqnarray} 
where $\mu = m_A \hbar^{-1}$, $\kappa_1=m_1 \hbar^{-1}$ and $\kappa_2 
=m_2 \hbar^{-1}$. 
In scalar $\varphi^4$ theory the propagator contains an $\hbar$ 
factor (like 
that in $iD_{\mn}$) which (with the $\hbar^{-1}$ at each vertex 
included) 
leads to the ``loop expansion'' in powers of $\hbar$ [37,38]. In this 
model, 
however, the absence of an $\hbar$ factor in $iS_F$ together with an 
$\hbar^{-
1}$ at each vertex causes loop integrals to contain factors 
$(\hbar^{-1})^n$, 
and it turns out that in Case B we must use $w=\hbar^{-1}$ as the 
expansion parameter. 
(We are not concerned with the classical limit $\hbar\to 0$, since 
we put 
$\hbar=1$ on completion of the renormalisation in Case B and can put 
$\hbar=1$ 
in Case A1 or A2 immediately.)

We refer to the $(k^2 -\mu^2)$ denominator and the pole at 
$k^2 = \mu^2$ in (13) as the primary denominator and primary pole 
of $D_{\mn}$, and refer to the analogous objects in the 
improper and renormalised propagators $i\cald_{\mn}^{}$, 
$i\cald_{\mn}^R$ by the same terms.

The topologies of the Feynman diagrams generated by $\hbar^{-1}\call_1$ 
are essentially the same as those of QED with two fermions, and, with the 
propagators (13), (14), the usual power-counting analysis [3,28] 
shows that the degree of divergence of an arbitrary  
diagram is the same as that of the analogous diagram in QED. The 
model is renormalisable.  However, our procedure 
differs from the usual complete cancellation of divergences order by 
order [39].

\section{Renormalisation of the Boson Propagator} 

The full boson propagator is 
\begin{eqnarray} i{\cal D}^{}_{\mn} 
= i { D}^{}_{\mn} +i { D}^{}_{\mu\sigma} [i\bar \pi^{\sigma\rho}] 
i {D}^{}_{\rho\nu} +\cdots \end{eqnarray} 
where 
$\bar\pi^{\sigma\rho}$, generated by $\hbar^{-1}\call_1$, is defined 
by \begin{eqnarray} 
\bar\pi^{}_{\mn}(k) = \pi^{}_{\mn}(k) -c^{}_3w(k^{}_\mu 
k^{}_\nu -g^{}_{\mn}k^2) +c_4w\xi^{-1} k_\mu k_\nu - 
m^2_A w^{3} g^{}_{\mn} , \end{eqnarray} where
$w=\hbar^{-1}$ and $\pi^{}_{\mn}(k)$ is the self-energy tensor (5). 
For the lowest 
order $\psi$-loop and $\eta$-loop components we find, using (14), 
that $\rho(k^2) \ne 0$ if $f\ne 0$ (the divergent part is given by 
(9)), and that $\tau(k^2) =0$ (as in QED). 

In Case A1 or A2, each of $m_A$ (by (6), (7)) and, it turns out 
below, $\pi_{\mn}$, 
$c_3$ and $c_4$ are $O(g^j)$, $j\ge 1$, and we can sum the series (15)
correctly to any given $O(g^n)$, in the usual way. In Case B the 
expansion parameter is $w$. It turns out that each term in (16) is of 
$O(w^j)$, $j\ge 1$, so that we can sum (15) correctly to any given 
$O(w^n)$. In every case we obtain
\begin{eqnarray} 
i{\cal D}^{}_{\mn}(k) = {-iw^{-1}\over [1+w^{-1} \pi(k^2) -
c^{}_3]k^2 -w^{2}m_A^2 -w^{-1}[\rho(k^2)-w^{3}m_A^2]} [g^{}_{\mn} 
+k^{}_\mu 
k^{}_\nu Q(k^2)], \end{eqnarray} 
where 
\begin{eqnarray} Q(k^2) = 
{\xi [1+ w^{-1}\pi(k^2) - c^{}_3 + w^{-1}\tau(k^2)] +c^{}_4 -1\over 
k^2 [1-\xi w^{-1} \tau(k^2)-c^{}_4] -\xi w^{-1}\rho(k^2)} ,  
\end{eqnarray} 
correctly to $O(g^n)$ in Case A1 or A2, or $O(w^n)$ in Case B, for 
any given $n$. 
We see the expected cancellation in the primary denominator 
of $\cal D_{\mn}$ between the $-m_A^2$, $m_A^2$ terms coming from 
$\call_0$ via (13), and $\call_1$.  
In equations holding in all cases, or in A1 or A2, we usually write 
$\hbar$ rather than $w^{-1}$ in what follows, while in Case B we 
usually write $\hbar$ as $w^{-1}$. The primary denominator 
of $\cald_{\mn}$ is, from (17),
\begin{eqnarray} d(k^2)= [1+\hbar\pi(k^2) - c^{}_3] k^2 -
\hbar\rho(k^2).  \end{eqnarray} 

We renormalise so as to place the zero of $d(k^2)$ at $k^2 = 
\calm^2= m_{AR}^2\hbar^{-2}$,
where $m_{AR}$ is the renormalised boson mass, which may be complex 
if the boson is unstable. We write
\begin{eqnarray} \calm^2 = M^2 (1-i\delta) \end{eqnarray}
and expand $d(k^2)$ about $k^2=\calm^2$, to obtain 
\begin{eqnarray} d(k^2) = e +\Omega^{-1}_3(k^2 -\calm^2) +R(k^2), 
\end{eqnarray} 
where 
\begin{eqnarray} e &=& [1+\hbar\hat\pi - c_3^{}] \calm^2 -
\hbar\hat\rho, \\ 
\Omega^{-1}_3 &=& 1+\hbar\hat\pi - c_3 
+\calm^2 \hbar\hat\pi' -\hbar\hat\rho' \end{eqnarray} 
and 
\begin{eqnarray} R(k^2) 
&=& \frac{\hbar}2 (2\hat\pi' +\calm^2\hat\pi''-
\hat\rho'') (k^2-\calm^2)^2  \nonumber \\
&&\quad +\frac{\hbar}6 (3\hat\pi'' 
+\calm^2\hat\pi''' -\hat\rho''') (k^2-\calm^2)^3
+\cdots, \end{eqnarray} 
where $\hat\pi=\pi(\calm^2)$, 
$\hat\rho = \rho(\calm^2)$, $\hat\pi',\hat\rho',\ldots$ are 
defined by (A.12), (A.13) in Appendix~A. $\Omega_3$ 
is the renormalisation factor multiplying the renormalised 
propagator $i\cald_{\mn}^R(k)$ (the analogous factor in QED is 
commonly written as $Z_3$).  

From (5) we see that $k^2 \pi(k^2)$, $\rho(k^2)$ are of the same 
dimension, and with $c=1$ and $\hbar$ dimensionless, $k^2$ has the 
dimension of $m^2$, so that we can write
\begin{eqnarray} \rho(k^2) = \sum m_i m_j \sigma^{ij}(k^2) 
\end{eqnarray}
with $\sigma^{ij}(k^2)$, $\pi(k^2)$ of the same dimension and with 
their components having the same $w$ and $g$ dependence (see Appendix 
A). $m_i$, $m_j$ stand for $m_1$, $m_2$, $m_A$ in all 
combinations; at least the $m_1^2$, $m_2^2$ terms are not zero. 

\subsection{Real mass, $\calm = M$} To obtain a renormalised real 
mass $m_{AR} = \hbar M$ we 
proceed loop by loop, in the $n$-loop and $(n+)$-loop sets of 
diagrams defined and discussed in Appendix A. We use the notation and 
results of Appendix~A with minimal further comment. Each $n$-loop 
set 
contains diagrams with $2n$ vertices but no counterterm or mass 
insertions, plus other diagrams of the same order 
containing counterterm vertex insertions (as usual) but also mass 
vertex insertions. We describe $(n+)$-loops, only present in Case 
A1, below. For the boson self-energy, each set gives rise to 
components of $\pi_{\mn}(k)$, i.e. of $\pi(k^2)$, $\tau(k^2)$, 
$\rho(k^2)$.  

For Case A2, the $n$-loop component of $\pi(k^2)$ is given by 
\[ \pi_{2n}^{(2)}(k^2) = g^{2n} 
[\pi_{2n,n}^{(2)}(\hbar)\eps^{-n} + \pi^{(2)}_{2n,n-1} 
(k^2,\hbar)\eps^{-n+1} +\cdots + \pi_{2n,0}^{(2)} (k^2,\hbar) \eps^0]
\quad{\rm(A.2)} \]
where $\pi^{(2)}_{2n,n}$ is 
independent of $k^2$ and real. There are 
parallel 
expressions for $\tau^{(2)}_{2n}(k^2)$ and 
$\sigma^{ij(2)}_{2n}(k^2)$. We define the following leading 
divergent parts: 
\[ \pi^{(2)}_{2nL} = g^{2n} 
\pi_{2n,n}^{(2)}\eps^{-n}, \quad \tau_{2nL}^{(2)} = g^{2n} 
\tau_{2n,n}^{(2)} \eps^{-n},\quad \sigma_{2nL}^{ij(2)} = g^{2n} 
\sigma_{2n,n}^{ij(2)}\eps^{-n}, \quad {\rm(A.4)} \]
 which 
are real and independent of $k^2$, so that (see (A.14)) 
\begin{equation} \hat\pi_{2nL}^{(2)} = \pi_{2nL}^{(2)}, \quad 
\hat\tau_{2nL}^{(2)} = \tau_{2nL}^{(2)}, \quad 
\hat\sigma_{2nL}^{ij(2)} = \sigma_{2nL}^{ij(2)} .  \end{equation}  

For Case B we have the components  
\[ \pi_{n,n}^{(B)}(k^2) = w^{n+1} 
[\pi_{n,n}^{(B)}\eps^{-n} + \pi^{(B)}_{n,n-1} (k^2)\eps^{-n+1} 
+\cdots],  \quad\qquad\quad\qquad {\rm(A.5)} \]
(where $\pi_{n,j}^{(B)}$ 
are functions of $g^2$) with similar expressions for 
$\tau_{n,n}^{(B)}(k^2)$ and $\sigma_{n,n}^{ij(B)}$, and leading 
divergent parts, independent of $k^2$ and real, analogous to 
those of Case A2 in (A.4); also, analogues of (26) hold.

For Case A1, the $n$-loop diagrams, each of which must contain an 
even number of fermion mass insertions, lead to components 
$\pi_{2n}^{(1)}(k^2)$, $\tau_{2n}^{(1)}(k^2)$, 
$\sigma_{2n}^{ij(1)}(k^2)$, and leading divergent parts, that are 
similar in form to those of their counterparts in Case A2, 
exemplified by (A.3), (A.4). In addition, there are diagrams 
containing an odd number of fermion mass insertions that generate 
terms of $O(g^{2n+1})$, for which there are no corresponding 
$2n$-vertex-only loops. We refer to such diagrams as $(n+)$-loop 
diagrams. They give the components 
\[ \pi_{2n+1}^{(1)}(k^2) = g^{2n+1} [ 
\pi_{2n+1,n}^{(1)} \eps^{-n} + \pi_{2n+1,n-1}^{(1)} (k^2)\eps^{-
n+1} +\cdots], \quad\qquad\qquad{\rm(A.7)} \]
where $n\ge 1$ and 
$\pi_{2n+1,n}^{(1)}$ is real and independent of $k^2$, plus 
components $\tau_{2n+1}^{(1)}(k^2)$ and 
$\sigma_{2n+1}^{ij(1)}(k^2) $ of similar form. Since each $(n+)$-
loop diagram may be regarded as a Case A2 $n$-loop into which an 
additional mass vertex has been inserted, and such an insertion 
cannot increase the degree of divergence of each of the $n$-loop 
components ($\pi_{2n}^{(2)}(k^2)$, etc.), we 
see that $\pi^{(1)}_{2n+1,n}$ could be zero, and similarly for 
$\tau,\rho$. The 
leading divergent parts of the components of $\pi(k^2)$, which 
are real and independent of $k^2$, are 
\[
\pi_{2nL}^{(1)} = 
g^{2n} \pi_{2n,n}^{(1)}\eps^{-n},\quad \pi_{(2n+1)L}^{(1)} = 
g^{2n+1}\pi^{(1)}_{2n+1,n}\eps^{-n} ,
\qquad\qquad\quad {\rm(A.8)} \]
where $\pi_{(2n+1)L}^{(1)}$ might be zero, and analogues of (26) 
hold.  

To renormalise loop by loop, we impose the condition 
\begin{equation} (\hbar \hat\pi_{qL}^{(J)} -  c_{3q}^{(J)}) M^2 - 
\hbar \sum m_i m_j \hat\sigma_{qL}^{ij(J)} =0, \end{equation} 
where the 
latter term is $\hbar \hat\rho_{qL}^{(J)}$, $J=1,2,B$ labels 
the case, and $q=2n$ for $J=2$, $q=n$ for $J=B$, and $q=2n,2n+1$ for
$J=1$.
The quantities in (27) are all independent of $k^2$. Since $g$ 
in Cases A1, A2, and $w=\hbar^{-1}$ in Case B, are variable expansion 
parameters, and $m_i,m_j$ are given by (6), (7), (8), we must 
have 
\begin{equation} M=Bg \ \mbox{(Case A1)}, \quad M=B g^2 \ 
\mbox{(Case A2)}, \quad M= B\hbar^{-1} = B w \ \mbox{(Case 
B)}. \end{equation} 
With $\hbar =1$ finally, our three cases have 
fermion and boson masses of $O(1)$, $O(g)$ and $O(g^2)$.  

The condition (27) determines $c_3$. We take the counterterm 
parameter $c_{3q}^{(J)}$ to have the form 
of $\hbar^{-1} \pi_{qL}^{(J)}$, and all other potential components of 
$c_3$ to be zero. In contrast to the usual complete cancellation 
of all divergences in QED and QCD (e.g.\ with minimal 
subtraction, MS), we effect only an order by order cancellation 
of the most divergent part of each component of $\pi(M^2)M^2 -
\rho(M^2)$, and do {\it not\/} make any cancellation of less 
divergent 
quantities.  

In Case A2 at one loop, we have, from (22), (27), \bq e_1 &=& 
(1+\hbar\hat\pi_{2L}^{(2)} -c_{3(2)}^{(2)}) M^2 -
\hbar\hat\rho_{2L}^{(2)} +O(\eps^0) \\ &=& O(\eps^0), \eq 
while, from (23), (27), 
\bq (\Omega_3^{-1})^{}_1 &=& 1+\hbar 
\hat\pi_{2L}^{(2)} - c_{3(2)}^{(2)} +O(\eps^0)\\ 
&=& O(\eps^{-1}), \eq 
and, since $\hat\pi_2^{(2)'}$, $\hat\rho_2^{(2)'}$, etc.\ are 
$O(\eps^0)$, \begin{equation}R_1(k^2) = O(\eps^0). \ee 
Accordingly, to one loop in Case A2 we have 
\bq \Omega_3 e = O(\eps),\\ 
\Omega_3 R(k^2) = O(\eps) \eq and, from (22), 
\begin{equation} d=\Omega_3^{-1} [k^2 -M^2 +O(\eps)]. \ee 
In Appendix B.1 we show that going to two loops in Case A2 gives (36) 
again. The result extends to $N$ loops, $N$ arbitrary. 

In Case B at one loop, with (22), (23), (27) and since 
$\hat\pi^{(B)}_{1L}$ is $O(w^2 g^2 \epsilon^{-1})$ and 
$\hat\rho_{1L}^{(B)}$ is $O(w^4 g^2 \epsilon^{-1})$, we see that 
\begin{eqnarray}
e_1 = (1+w^{-1} \pi_{1L}^{(B)} -c_3^{(B)}) M^2 -w^{-1} 
\rho_{1L}^{(B)} =O(\epsilon), \end{eqnarray}
and $\Omega^{-1}$ is $O(\eps^{-1})$, so that we obtain (34), (35), 
(36) again. The result extends to $N$ loops. In Appendix B.1 we 
obtain (36) in Case A1, proceeding from 1-loop to (1+)-loops, 
2-loops, (2+)-loops and so on. 

We see from (17), (19) and (36) that $\Omega_3$ is the 
renormalisation factor that multiplies the 
renormalised propagator $i\cald_{\mn}^R$ obtained below. We also 
see, using (A.3), (A.5), (A.7), that $(\Omega^{-1}_3)^{}_n$ and 
$(\Omega_3^{-1})^{}_{n+}$ are $O(\eps^{-n})$, while $R_n(k^2)$, 
$R_{(n+)}(k^2)$ are $O(\eps^{-n+1})$.  

The BRS Ward identities of (1) impose no constraint on $Z_4$ (the
$\xi$ dependence being unphysical), allowing $c_4$ to be chosen 
arbitrarily.
To renormalise $Q(k^2)$, we take the components of $c_4$ to be 
\begin{equation} 
c_{4q}^{(J)} = -\xi [\lambda^{-1} (\hbar 
\pi_{qL}^{(J)} - c_{3q}^{(J)}) +\hbar\tau_{qL}^{(J)} ], \ee 
where 
$\lambda$ is arbitrary. Working up to $N$-loops or $(N+)$-loops, 
we see that in each case the numerator and denominator of 
$Q(k^2)$  are dominated, as $\eps\to 0$, by the $O(\eps^{-N})$ 
terms in $\pi_{UL}^{(J)}$, $c_{3U}^{(J)}$, $\tau_{UL}^{(J)}$, 
$\sigma_{UL}^{ij(J)}$, where $U=2N,\, 2N+1$. Using (27), we obtain 
\bq Q(k^2)_U^{(J)} 
&=& \frac{ (1-\lambda^{-1}) 
(\hbar\pi_{UL}^{(J)} - c_{3U}^{(J)} )} { (k^2\lambda^{-1} -M^2) 
(\hbar \pi_{UL}^{(J)} -c_{3U}^{(J)} )} +O(\eps)\nonumber \\
 &=& \frac {\lambda-1 } {k^2 -\lambda M^2} +O(\eps), \eq 
which, as $\eps 
\to 0$, is independent of $J$, $N$ and $\xi$. The gauge parameter 
$\xi$ has been replaced by an unrelated, arbitrary parameter 
$\lambda$, which we refer to as the pseudo-gauge parameter.

From (17), (19), (23), (36) and (39) we obtain, as $\eps\to 
0$, the renormalised propagator 
\begin{equation} 
i\cald_{\mn}^{R(loop)}(k) 
= \frac {-i\hbar}{k^2 -M^2 +i\eps'} \left[ g_{\mn} +(\lambda-1) 
\frac{ k_\mu k_\nu}{k^2 -\lambda M^2 +i\eps'} \right] \ee (where 
we have re-inserted $i\eps'$) and the renormalisation factor 
\begin{equation} \Omega_3 = [1+\hbar \hat\pi - c_3 + M^2 
\hbar\hat\pi' -\hbar\hat\rho']^{-1}, \ee 
which are correct to 
arbitrary order. The propagator (40) is independent of~$\xi$. 

\subsection {Complex mass, $m_{AR} =\hbar \calm$} For a complex mass
$\hbar\calm$ we 
cannot renormalise loop by loop, since (27) cannot hold if $M^2$ 
is replaced by $\calm^2$. Instead, we renormalise order by order. 
With $\calm^2$ of the form (20) and $M$ given by (28), we take, for 
simplicity, $\delta$ to be of $O(g^0 \hbar^0)$, so that $\calm^2$ 
is $O(g^2)$, $O(g^4)$ or $O(w^2)$ in Cases A1, A2 or B. The 
treatment that follows can be modified to accommodate $\delta = 
\delta_0 +\delta_1 g +\cdots$ in A1 or A2 (at any given order 
$O(g^2)$, a term $\delta_\sigma g^\sigma$ leads to $O(g^n)$ terms 
that are less divergent than the dominant divergent 
components) or $\delta = \delta_0 +\delta_1w +\cdots $ in 
Case~B.  

Again we determine $c_3$ by (27), and (28) holds.
Up to $O(g^2)$ in A1, A2 and $O(w)$ in B, 
we have \be 
 e_{g2}^{(1)} = B^2 g^2 (1-i\delta), \quad e_{g2}^{(2)} =0, \quad
e_w^{(B)} =0, \ee 
which are $O(\eps^0)$ (there is no need to 
differentiate B into B$^{(1)}$, B$^{(2)}$, B$^{(B)}$, or 
$\delta$, $\beta_1,\ \beta_2,\ \beta_A$ similarly, in what 
follows), and, with $\hbar=1$ in A1, A2, 
\bq (\Omega_3^{-1})_{g2}^{(1)} &=& 1+g^2 \tilde 
\pi_{2,1}^{(1)} \eps^{-1} +O(\eps^0),\\ 
(\Omega_3^{-1})_{g2}^{(2)} 
&=& 1+g^2 \tilde \pi_{2,1}^{(2)} \eps^{-1} +O(\eps^0),\\ 
(\Omega_3^{-1})_{\hbar}^{(B)} &=& 1+w \tilde \pi_{1,1}^{(B)} 
\eps^{-1} +O(\eps^0), \eq 
which are $O(\eps^{-1})$. Here we have written 
\be \hat\pi_{q,n}^{(J)} -\hbar^{-1} c_{3(q)}^{(J)} = 
\tilde\pi_{q,n}^{(J)}.  \ee 
In addition, we see from (24) that 
$R_{g2}^{(1)}(k^2)$, $R_{g2}^{(2)}(k^2)$ and 
$R_{\hbar}^{(B)}(k^2)$ are each $O(\eps^0)$. Consequently, we 
obtain (34), (35) and (36) to $O(g^2)$, $O(w)$ in Cases A1, 
A2 and B.  

In Appendix B we continue the argument to $O(g^4)$, $O(g^6)$ in Case 
A2, $O(w^2)$, $O(w^3)$ in Case B and $O(g^3)$, $O(g^4)$ in Case A1.
It is evident that we can continue to an arbitrarily high order. 
The procedure
``cuts across loops''. At lowest order 
it links the  ``zero loop'' $\calm^2$ to a one-loop term. Up to 
$O(g^3)$ in Case A1, it involves zero-loop, one-simple-loop and 
one-loop-with-mass-insertion contributions. In A2 up to $O(g^4)$ 
and B up to $O(w^2)$, it links zero-loop and two-loop terms, 
and up to $O(g^6)$, $O(w^3)$ links one-loop and three-loop 
contributions.  

The similarity in the renormalisation of Cases A2 and B shows 
that, in a sense, $w$ is equivalent to $g^2$. If we had 
chosen the masses $m_1$, $m_2$, $m_A$ in Case B to be of the form 
$m=\beta w^{1/2}$, then renormalisation in Case B would 
proceed as it does in Case A1, i.e.\ up to $w, 
w^{3/2},$ ... etc.  

With the components of $c_4$ given by (38), we expand the 
numerator and denominator of $Q(k^2)$ up to $O(g^{2N})$ or 
$O(g^{2N+1})$, in Cases A1 and A2, and to $O(w^N)$ in Case B, 
with $N$ arbitrarily large. The highest divergence, of $O(\eps^{-
N})$, is in $\pi(k^2)$, $\hat\pi_L$, $c_3$ to these orders, while 
the highest divergence in $\rho$ to these orders is $O(\eps^{-
N+k})$, because of the factors $m_i m_j$, which are of $O(g^2)$, 
$O(g^4)$, $O(w^2)$ in the three cases, in $\rho$. 
Consequently, to $O(g^{2N})$, $O(g^{2N+1})$, $O(w^N)$, $Q$ is 
\be Q(k^2) = \frac{ (1-\lambda^{-1}) 
\tilde\pi_{UN}^{(J)}[1+O(\eps)]} {k^2 \lambda^{-1} 
\tilde\pi_{UN}^{(J)} [1+O(\eps)]} \ee 
where $U=2N,2N+1$. In the 
limit $\eps\to 0 $ we obtain (cf.\ (39))
\be 
Q(k^2) = \frac{\lambda-1}{k^2}, 
\ee which is independent of $N$ and $\xi$, and the renormalised 
propagator is
\be 
i\cald_{\mn}^{R(order)}(k) 
= \frac{-i\hbar}{k^2 -\calm^2 +i\eps'} \left[ g_{\mn} +(\lambda-1) 
\frac{k_\mu k_\nu}{k^2 +i\eps'} \right] .\ee 
The renormalisation 
factor $\Omega_3$ is again given by (41).

\subsection{Choice of $\lambda$}
The propagators (40), (49) contain unphysical poles at $k^2=\lambda M^2$,
$k^2=0$, as does the unrenormalised propagator (13) at $k^2=\xi\mu^2$.
For a stable boson $(\calm=M < 2m_1)$, we may take the limit $\lambda
\rightarrow\infty$ to obtain
\be
i\cald_{\mn}^{(R)(\infty )}(k) = 
\frac{-i\hbar}{k^2 -M^2 +i\eps'} 
\left[ g_{\mn} -\frac{k_\mu k_\nu}{M^2} \right] ,
\ee
which is the form usually expected for a massive vector boson.  In theories
with SSB (or in (13)), this propagator would result from unitary gauge
$(\xi\rightarrow\infty )$, but here we can take $\lambda\rightarrow\infty$
for any value of $\xi$, the unitary pseudo-gauge.
However, we cannot take $\lambda\rightarrow\infty$ in (49) for an
unstable boson.

The choice of $\lambda =1$ in $c_4$ results in the same propagator for
either stable or unstable bosons, viz.,
\be
i\cald_{\mn}^{R(1)}(k) =
\frac{-i\hbar\cdot g_{\mn}}{k^2-\calm^2 +i\eps'} .
\ee
There is no discontinuity in form passing from $M < 2m_1$ to $M > 2m_1$.
We can choose $\lambda =1$ for any value of $\xi$.  In a sense, $\xi$ in
(13) has been renormalised to $\lambda$, $\lambda =1$, or in (50), for
a stable boson only, to $\lambda$, $\lambda\rightarrow\infty$.

The Ward identities with non-zero boson mass imply that it is the spin-0 
part of the virtual $f\bar{f}$ self-energy loop that acts as an effective 
longitudinal degree of freedom [40].

\section{Renormalisation of the Fermion Propagators}

The full improper propagator for the $\psi$ or $\eta$ fermion is 
given by the series
\be
i \calS_F(\slnp) = i(\slnp -\kappa) (1+u +u^2 +\cdots), \ee
where
\be
u = [\Sigma(\slnp) - \kappa +s\slnp](\slnp -\kappa)^{-1},
\ee
in which $\Sigma(\slnp)$ is given by the self-energy $-
i\Sigma(\slnp)$, the first $\kappa =m\hbar^{-1}$ is from $-\hbar^{-
1}\chi$ in $\hbar^{-1}\call_1$, and $s$ is a counterterm parameter 
given below. Here $m$, $\Sigma$, $\kappa$, $s$ are $m_j$, $\Sigma_j$, 
$\kappa_j$, $s_j$, with $j=1$ for $\psi$, $j=2$ for $\eta$. We 
often omit these subscripts in what follows. We write
\be
\Sigma(\slnp) = -a(p^2)(\slnp) +\kappa b(p^2), \ee
where
\be a(p^2) = a_{(1)}(p^2) +a_{(5)}(p^2)\gamma^5,\qquad
b(p^2) = b_{(1)}(p^2) +b_{(5)}(p^2)\gamma^5 \ee
(we note that $b_{(5)}(p^2) =0$ at one loop), more explicitly, we 
have $a_{j(1)},\ a_{j(5)}$, $b_{j(1)},\ b_{j(5)}$. Similarly we write
\be s = s_{(1)} +s_{(5)}\gamma^5, \ee
where, from (1),
\bq
s_{1(1)} &=& {\tes\frac12}(c_{2L} +c_{2R}), \qquad 
s_{1(5)} = {\tes\frac12} (c_{2L}-c_{2R}),\\
s_{2(1)} &=& {\tes\frac12} (c_{6L} +c_{6R}), \qquad 
s_{2(5)} = {\tes\frac12} (c_{6L}-c_{6R}).\eq

We take
\be f \ne \pm 1,0,\ee
$f\ne \pm1$ so that at one loop the mass-regenerating terms 
$b_1(p^2)$, $b_2(p^2)$ do not vanish, and 
$f\ne0$ so that at one loop the boson mass (section~3) does not 
vanish.

In Case A1 or A2 we can sum (52) correctly to any given $O(g^n)$, and 
in Case B, 
correctly to any given $O(w^n)$ (see the discussion 
of boson propagator series summation in section~3), to obtain, using (53), 
(54),
\bq
i\calS_F(\slnp)
&=& i(\slnp -\kappa)^{-1} (1-u)^{-1} \nonumber\\
&=& i [(1-u)(\slnp-\kappa)]^{-1}\\
&=& i [\slnp -\kappa -(\Sigma -\kappa +s\slnp)]^{-1}\\
&=& i \{ [1+ a(p^2) -s] \slnp -\kappa b(p^2)\}^{-1},\eq
where we have used $A^{-1} B^{-1} =(BA)^{-1}$ to reach (60) ($u$ 
contains $\gamma^5$). The 
expected cancellation of $-\kappa$, $\kappa$ from $\call_0$, 
$\call_1$ may be seen in (61); however, the term $\kappa$ in the 
propagator $i (\slnp-\kappa)^{-1}$  has been responsible for the 
generation of the mass term $\kappa b(p^2)$, which leads below to the 
renormalised mass $m$. 

We expand $a(p^2)$ about $\slnp = \kappa$ to obtain
\bq
a(p^2) &=& \hat a +(p^2 -\kappa^2) \hat a' +{\tes\frac12} (p^2-
\kappa^2)^2 \,\hat a'' +\cdots
\nonumber\\
&=& \hat a +2\kappa \hat a' (\slnp-\kappa) +F(\slnp)(\slnp-
\kappa)^2\,,\eq
and expand $b(p^2)$ similarly; where $\hat a,\hat b,\hat a',\ldots$ 
are defined 
in Appendix A (following (A.13)) and $F(\slnp)$ depends on observing 
the 
order of the factors shown in (63). We expand about $\slnp =\kappa$, 
i.e.\ about 
$\slnp =\kappa_1$, $\slnp=\kappa_2$, because the renormalised masses 
$m_{1R}$, $m_{2R}$ must 
equal the initial masses $m_1$, $m_2$, since each fermion is stable, 
so that the 
$S$-matrix reduction formula must operate in the usual way.
Using (63), the denominator
of $i\calS_F(\slnp)$ in (62) becomes
\bq
d_f &=& [1+ a(p^2) -s]\slnp - \kappa b(p^2) \nonumber\\
&=& E +\Omega_2^{-1}(\slnp-\kappa) + T(\slnp)(\slnp-\kappa)^2 \eq
where, with $\hat a = a(\kappa^2)$, $\hat b = b(\kappa^2)$, 
\be
E = (1+\hat a-s -\hat b)\kappa,\ee
\be
\Omega_2 = [ 1+\hat a-s +2\kappa^2 (\hat a' -\hat b') ] ^{-1} \ee
(the factor analogous to $\Omega_2$ in QED is commonly written as 
$Z_2$).

We renormalise loop by loop, as was done in section 3.1 for the 
boson.
In Cases A1, A2 and B the leading divergent parts of $\hat a$ are, 
from (A.10) and the analogues of (A.14),
\bq \left.
\begin{array}{l}
\hat a_{j,2nL}^{(1)} = g^{2n} a^{(1)}_{j,2n,n} \epsilon^{-n},\qquad
\hat a_{j,(2n+1)L}^{(1)} = g^{2n+1} a^{(1)}_{j,2n+1,n} \epsilon^{-
n},\\
\hat a_{j,2nL}^{(2)} = g^{2n} a^{(2)}_{j,2n,n} \epsilon^{-n},\qquad
\hat a_{j,nL}^{(B)} = w^{n+1} a^{(B)}_{j,n,n} \epsilon^{-
n},\end{array} \qquad \right\}
\eq
and similarly for $\hat b$. It is convenient to introduce the 
notation, similar to that in (27), of $\hat a_{jqL}^{(J)}$, $\hat 
b_{jqL}^{(J)}$, where $J=1,2,B$ labels 
Cases A1, A2, B, and $q=2n$ for $J=2,B$, while $q=2n,2n+1$ for
$J=1$. From Appendix A (see (A.9), (A.12), (A.13)) we see that 
$\hat a^{'(J)}_{jq}, \hat b^{'(J)}_{jq}, \hat a^{''(J)}_{jq}, 
\hat b^{''(J)}_{jq},\ldots$ are $O(\epsilon^{-n+1})$. 
From (63) and the similar expansion of $b(p^2)$, and (65), (66), it 
follows that
 $T(\slnp)$ only involves $\hat a', \hat b',\ldots$, so that, up to 
$n,(n+)$-loops,
\be T(\slnp)_{n(n+)} = O(\epsilon^{-n+1}). \ee

We impose the condition (cf. (27)) that
\be \hat a_{jqL}^{(J)} - s_{jq}^{(J)} - b^{(J)} _{jqL}=0,\ee
which is to hold order by order, and determines the
components of the counterterm parameter $s$, viz.,
\be
s_{jq}^{(J)} = g^q (a_{j,q,n}^{(J)} - b_{j,q,n}^{(J)} ) \epsilon^{-
n}. \ee
All other potential components of $\kappa$ are taken to be zero. We 
write
\be s_{jq}^{(J)} = s_{jq(1)}^{(J)} +s_{jq(5)}^{(J)} \gamma^5. \ee
The counterterm parameters $c_{2L}^{(J)}$, $c_{2R}^{(J)}$, 
$c_{6L}^{(J)}$, 
$c_{6R}^{(J)}$ are then given by (57), (58), (70) and (71). 

A discussion like that in section 3.1 for the loop by loop 
renormalisation of 
the boson propagator and the use of (64), (65), (66), (68) and (69) 
shows that, 
up to $n(n+)$-loops,
\be
E = O(\epsilon^{-n+1}), \qquad \Omega_2^{-1} = O(\epsilon^{-n}), \ee
\be
\Omega_2E = O(\epsilon),\qquad \Omega_2 T(\slnp) = O(\epsilon), \ee
so that (64) gives
\be
d_{fj} = \Omega_{2j}^{-1} [\slnp -\kappa_j +O(\epsilon)] \ee
for the denominators of the $\psi$ and $\eta$ propagators, with
\be
\Omega_{2j}^{(J)} = [ 1+ \hat a_j^{(J)} -s_j ^{(J)} +2\kappa_j^2 
(\hat a_j^{'(J)} -\hat b_j^{'(J)})] ^{-1},
\ee
which contains $\gamma^5$. From (62), (64) and (74), we obtain, using 
$(AB)^{-1} =B^{-1}A^{-1}$,
\be
i\calS_{Fj}(\slnp) = \frac{i}{\slnp -\kappa_j +i\epsilon' 
+O(\epsilon)} \cdot
\Omega_{2j}^{(J)} \ \ . \ee
In the limit $\epsilon\to 0$, we obtain the renormalised propagators 
\be
i\calS_{F\psi}^R (\slnp) = \frac{i}{\slnp -\kappa_1 +i\epsilon'}\ , 
\qquad
i\calS_{F\eta}^R (\slnp) = \frac{i}{\slnp -\kappa_2 +i\epsilon'} \ . 
\ee
The renormalisation factors $\Omega_{2\psi}^{(J)} = 
\Omega_{21}^{(J)}$, $\Omega_{2\eta}^{(J)} = \Omega_{22}^{(J)}$ stand 
to the right of these propagators.
We can interpret the result that $\Omega_{2\psi}^{(J)}$, 
$\Omega_{2\eta}^{(J)}$ 
contain $\gamma^5$ by saying that, similarly to the situation in 
standard electroweak theory [37], the left and right components of 
$\psi,\eta$ are 
renormalised differently, since, dropping $j$, $J$, we can write
$\Omega_2 = \tilde\Omega_2 (1+\Lambda\gamma^5)$, with 
$\tilde\Omega_2,\Lambda$ free of $\gamma^5$, so that (76) becomes, as 
$\epsilon\to 0$,
\be
i\calS_F^R(\slnp) = \tilde\Omega_2 \left[ (1+\Lambda) \frac i{\slnp-
\kappa+i\epsilon'} a_R
+(1-\Lambda) \frac i{\slnp-\kappa+i\epsilon'} a_L \right], \ee
where
$a_R = \frac12(1+\gamma^5)$, $a_L = \frac12(1-\gamma^5)$.

\section{The Vertices}

It is straightforward to write out the sequence of 
all possible skeleton diagrams, up to any given order $O(g^{2n})$, 
for a given 
matrix element $S_{f_i}$; with the definition that, for this model, a skeleton 
diagram contains no self-energy insertions and no vertex loops, which 
means no 
subgraphs with only two ``external'' lines and none, except for point 
vertices,
with three ``external'' lines. We then replace each propagator and 
point vertex
by the full improper propagator and full vertex part, each taken to an 
arbitrarily high order. We assume that each such propagator self-energy 
and vertex part comprises unambiguous integrals in $d=4-2\eps$ 
dimensions, and that we can 
assign the counterterms order by order in such a way as to effect the partial 
cancellation of the divergences in $\pi$ and $a(p^2)$ that we have 
made above, 
at each order $O(g^{2n})$ or $O(w^n)$. We gloss over the problem of overlapping 
divergences [38], treatable by standard methods [19].
The BPHZ renormalisation method [38] can be used to effect an explicit 
renormalisation of the model. 

Proceeding from these assumptions, we see that when the full 
propagators and 
vertex parts are inserted at each $\psi\psi A$ vertex  in a skeleton 
diagram, we obtain the product
\be P = 
\ldots i\calS_F^{(R)} \Omega_2 (-i g\hbar^{-1}) \gamma^\mu [Z_1 
+fZ_{1(5)}\gamma^5 +V(p,p')] i\calS_F^R (p') \Omega_2 \Omega_3^{1/2} 
i \cald_{\mn}^R(k)\ldots, \ee
in which the $Z_1$, $Z_{1(5)}$ terms come from (1) and $V$ is the 
vertex part 
$V(p,p') = V_{(1)} (p,p') +V_{(5)}(p,p') \gamma^5$. The interaction 
vertex is
\be
I_V = \Omega_3^{1/2} \Omega_2 \{ 1-c_1 +V_{(1)} (p,p') -[f(1-
c_{1(5)}) +V_{(5)}(p,p')]\gamma^5 \} (-ig\hbar^{-1}\gamma^\mu). \ee

We define the leading divergent parts
\bq
\Omega_{3L}^{-1} &=& 1+\hbar \tilde\pi_L- c_3 \\
\Omega_{2L}^{-1} &=& 1+ a_L-s \\
&=& (\Omega_{2L}^{-1})_{(1)} +(\Omega_{2L}^{-1})_{(5)}\gamma^5 , \eq
in which we have suppressed the $j,J$ labels carried by 
$\tilde\pi_L$, $a_L$ (of 
the forms given by (A.15), (A.16), $s$, $\Omega_{3L}^{-1}$ and 
$\Omega_{2L}^{-1}$. From (41), (75) and the analysis of orders $(g^n 
\epsilon^{-j})$ 
given in sections 3 and 4,
\be
\Omega_3^{-1} = \Omega_{3L}^{-1} [1+O(\epsilon)], \qquad
\Omega_2^{-1} = \Omega_{2L}^{-1} [1+O(\epsilon)]. \ee
Similarly we obtain
\be
V_{(1)}(p,p') = V_{L(1)}[1+O(\epsilon)], \qquad
V_{(5)}(p,p') = V_{L(5)}[1+O(\epsilon)], \ee
where the leading divergent parts $V_{L(1)}$, $V_{L(5)}$, of forms 
given by (A.19), (A.20), are independent of $(p,p')$ (we have 
suppressed $j,J$ labels).

We write (80) as
\be
I_V = \Omega_{3L}^{1/2} \Omega_{2L} (\alpha + \beta\gamma^5) (1-
f\gamma^5) [1+O(\epsilon)] (-ig\hbar^{-1}\gamma^\mu), \ee
where
\be
\alpha = (1-f^2)^{-1} \{ 1-c_1 +V_{L(1)} -f[f(1-c_{1(5)}) 
+V_{L(5)}]\}, \ee
\be
\beta =  (1-f^2)^{-1} \{ f(1-c_1 +V_{L(1)}) -[f(1-c_{1(5)}) 
+V_{L(5)}]\}. \ee
We choose $\alpha,\beta$ to be, i.e.\ we choose the values of 
$c_1,c_{1(5)}$ to be such that,
\be
\alpha = (\Omega_{2L}^{-1})_{(1)} \Omega_{3L}^{-1/2}, \qquad
\beta = (\Omega_{2L}^{-1})_{(5)} \Omega_{3L}^{-1/2}. \ee
Then (86) reduces to
\be
I_V = (1-f\gamma^5) [1+O(\epsilon)] (-ig\hbar^{-1}\gamma^\mu), \ee
so that, taking $\epsilon\to 0$, the renormalised $\psi\psi A$ 
coupling at vertices is 
\be
-ig_R\hbar^{-1}\Gamma^\mu = -i g\hbar^{-1}\gamma^\mu (1+f\gamma^5), 
\ee
with $g_R=g$,
which is the original unrenormalised coupling in $\call_{\rm eff}$, 
(1), divided by $\hbar$.

In Appendix C we find the values of $c_1,c_{1(5)}$ that give this 
result, and mention the similar treatments to obtain $c_1,c_{1(5)}$ 
in Cases B and A1.

In a similar way, we can obtain the values of $c_5,c_{5(5)}$ that 
renormalise the $\eta\eta A$ coupling to its original form, $+ig 
\bar\eta \gamma^\mu (1+f\gamma^5)\eta A_\mu$, in each of Cases A1, A2 
and B.

\section{Summary and Discussion}

We have demonstrated that a theory in which the action possesses chiral and 
gauge symmetries has a {\it perturbative\/} solution with nonzero gauge boson 
masses~(17), (19), (21), (40) and fermion masses~(62), (64), (77). The axial 
vector coupling $gf$ must be nonzero.  The essential mechanism is:
{\it the tree-level fermion masses of $\call_0$ cancel 
in $i{\cal S}_F$ but leave an indirect, non-vanishing effect in $\Sigma$~(54).}

A full Lagrangian density $\call$ does not by 
itself imply a unique spectrum.
Starting from the chiral- and BRS-invariant
$\call_{\rm eff}$, (1), we choose an $\call_0$ that contains a 
massive boson and massive fermions, then renormalise in a way that
dynamically regenerates via quantum corrections the same fermion masses
and the same boson mass or, in the unstable case, a complex boson mass. 
This choice of a chirally-asymmetric $\call_0$ and consequent asymmetric 
states is analogous to the choice of asymmetric rather 
than symmetric solutions in Schwinger-Dyson equations [9,10].
As usual in perturbation theory, our choice 
of $\call_0$ also breaks gauge symmetry: the partial-action $S_0 = 
\int d^4 x\call_0$ does not possess $U(1)$ BRS-invariance. However,
$S$-matrix elements are gauge-invariant. 

On renormalisation to one loop, the model contains a massive gauge boson
and two massive fermions, and we sketch a renormalisation procedure extending 
to any order. The renormalised propagators and couplings, and so $S$-matrix 
elements, are independent of the gauge parameter $\xi$. The 
model can be renormalised to all orders because the full Lagrangian
$\call_{\rm eff}$, (1), can be. No
new counterterms beyond the types exhibited in (1) (in particular, no
mass counterterms) should appear. In combination with causality and
dispersion relations,
renormalisability implies unitary high-energy behaviour.
A proof of renormalisability would make use
of the BRS Ward identities of (1), similar to the analogous identities of 
[9,40], with three important differences: our model is anomaly-free; it is 
solved perturbatively after the finite shift of masses from zero; and the
gauge and fermion masses are {\it independent} of one another.  This
independence is a result of their {\it separate} renormalisations and the
fact that the fermion self-energies are ``hard'' constant masses.  Jackiw and
Johnson, on the other hand, obtained an alternate solution with a ``soft''
fermion self-energy and finite relation between $m_{AR}$ and $m_{1,2R}.$

Some questions remain to explored. (1) The issue of unitarity with unstable 
particles is mentioned in section~1. (2)
A complete treatment also requires consideration of vacuum symmetries,
energy, and stability. (3) We have introduced a Dirac mass ansatz for the
fermions, but Majorana masses might also be possible. These would break
the residual $U(1)$ vector symmetry of fermions left after the full
chiral symmetry is broken. (4) The connection between our perturbative
treatment and previous non-perturbative solutions [7,8,9,10] is not 
fully elucidated.

The mechanism presented here is used in an 
electroweak theory that contains only $W$, $Z$, photon, ghost, lepton 
and quark fields, and in which renormalisation to one loop gives the 
particles their final masses [41].

\bigskip
\noindent{\bf Acknowledgments}

\bigskip
One of us (A.F.N.) wishes to thank Angas Hurst of The University of Adelaide 
for his encouragement. The other (D.C.K.) acknowledges helpful discussions with
Charles Thorn, Arthur Broyles, and John Klauder of the University of Florida,
and Roman Jackiw of M.I.T. 
This research was supported in part by the U.S.~Department of Energy under 
Grant No.~DE-FG05-86-ER40272.

\bigskip
\renewcommand{\theequation}{A.\arabic{equation}}
\setcounter{equation}{0}
\appendix{\bf Appendix A. Definitions and Properties of 
$\pi_{2n}(k^2)$, etc.}
\bigskip 

We have placed powers of $\hbar$ (dimensionless but not yet unity) in 
$\call_{\rm eff}$, (1), because $\hbar^{-1}$ is the expansion 
parameter in Case B, rather than $g$ as in Cases A1, A2. When $\hbar 
\ne 1$, the boson propagator 
(13) carries a factor $\hbar$, and diagrams are generated by 
$\hbar^{-1}\call_1$ [37,38], as illustrated by the 
calculation of 
$\cald_{\mn}(k)$ in section~3.

That set $S^{00}$ of diagrams which contribute to $\pi_{\mn}(k)$ and 
contain 
$2n$ fermion--fermion--boson ({\it ffb\/}) vertices, but no 
counterterm or mass insertion vertices, generates components 
$\pi_{2n}^{00}(k^2)$, 
$\tau_{2n}^{00}(k^2)$ and $\sigma_{2n}^{ij00}(k^2)$ (see (17)) which 
are of order $O(w^{n+1} g^{2n})$, as is easily seen by drawing 
diagrams. 
To the set $S^{00}$ we add, as usual, all diagrams containing fewer 
vertices but all possible combinations of counterterm insertions such 
that the order remains 
$O(w^{n+1}g^{2n})$, to obtain the set $S^0$. In line with the results 
of calculations beyond one loop in QED and QCD given in the 
literature and the discussion of a two-loop case given by 
Collins [19], we assume that $S^0$ generates a component of 
$\pi^0(k^2)$ of the form
\be
\pi^0_{2n}(k^2) =w^{n+1} g^{2n} [\alpha_{2n,n}\eps^{-n} 
+\alpha_{2n,n-1}(k^2) \eps^{-n+1} +\cdots+ \alpha_{2n,n}(k^2) ]
\ee
where $w=\hbar^{-1}$ and $\alpha_{2n,n}$ is independent of $k^2$ and 
real, plus components 
$\tau^0_{2n}(k^2)$, $\sigma_{2n}^{ii0}(k^2)$ of $\tau^0(k^2)$, 
$\sigma^{ij0}(k^2)$, of the same form. Similarly, we assume  that the 
$S^0$ sets 
for the fermion self-energies $\Sigma_1(\slnp)$, $\Sigma_2(\slnp)$ 
generate components of the form of 
\be 
a^0_{1(2n)}(p^2) = w^n g^{2n} [\beta_{2n,n}\eps^{-n} +\beta_{2n,n-
1}(p^2) \eps^{-n+1} + \cdots + \beta_{2n,n} (p^2)] 
\ee
for the functions $a_1^0(p^2)$, $b_1^0(p^2)$, 
$a_2^0(p^2)$, $b_2^0(p^2)$ (see section~4) in which the leading $w^n 
g^{2n} \eps^{-n}$ terms are again independent of $k^2$ and real. In 
the 
same way, we assume that the $S^0$ set for the vertex function 
$V(p,p') =V_1 
+V_5\gamma^5$, defined following (79), generates expressions similar 
to (A.2) for $V_1$ and $V_5$ but with $\beta_{2n,n-\sigma}(p^2)$ 
replaced by $\beta_{2n,n-\sigma}(p,p')$.   

For $\pi_{\mn}(k^2)$ in Case A2, (7), we add to $S^0$ all diagrams 
containing $(2n-j)$ {\it ffb\/} vertices, mass vertex insertions 
(each of $O(g^2)$) and 
counterterm insertions such that the order remains $O(g^{2n})$. We 
refer to the resulting set of diagrams, $S^{(2)}$, as the $n$-loop 
diagrams for Case A2. 
Adding a fermion or boson mass insertion (generated  by the $-
\hbar^{-1}\chi$ terms in $\hbar^{-1}\call_1$) to a diagram cannot 
increase its degree of
divergence. Using (A.1), it is then easy to see that $S^{(2)}$ 
generates the $n$-loop component 
\begin{eqnarray} 
\pi_{2n}^{(2)}(k^2) = g^{2n} 
[\pi_{2n,n}^{(2)}(w)\eps^{-n} + \pi^{(2)}_{2n,n-1} 
(k^2,w)\eps^{-n+1} +\cdots + \pi_{2n,0}^{(2)} (k^2,w)] 
\end{eqnarray}
in which $\pi^{(2)}_{2n,n}$ is real and independent of $k^2$; and 
similarly  
generates components $\tau_{2n}^{(2)}(k^2)$, 
$\sigma_{2n}^{ij(2)}(k^2)$. 
We suppress, in the notation, the dependence of the coefficients in 
(A.3), and in 
the parallel series for $\tau_{2n}^{(2)}$, $\sigma_{2n}^{ij(2)}$, on 
$w$ (in Case A2 and in Case A1 below, we may put $w=\hbar^{-1}=1$ 
already at this point). We write the 
real, $k^2$-independent leading divergent parts as
\begin{equation} 
\pi^{(2)}_{2nL} = g^{2n} 
\pi_{2n,n}^{(2)}\eps^{-n}, \quad \tau_{2nL}^{(2)} = g^{2n} 
\tau_{2n,n}^{(2)} \eps^{-n},\quad \sigma_{2nL}^{ij(2)} = g^{2n} 
\sigma_{2n,n}^{ij(2)}\eps^{-n}.
\end{equation} 

In Case B, we similarly add to $S^0$ all diagrams of {\it ffb\/} 
vertices plus mass and 
counterterm insertions such that the order is $O(w^{n+1})$, and refer 
to the 
resulting set of diagrams, $S^{(B)}$, as the $n$-loop diagrams for 
Case B. We see, using 
(A.1), that $S^{(B)}$ generates the $n$-loop component
\begin{eqnarray} \pi_{n,n}^{(B)}(k^2) = w^{n+1} 
[\pi_{n,n}^{(B)}\eps^{-n} + \pi^{(B)}_{n,n-1} (k^2)\eps^{-n+1} 
+\cdots],  \end{eqnarray}
where $\pi_{n,n}^{(B)}$ is real and independent of $k^2$, and we have 
suppressed (but only in the notation) the dependence of the 
coefficients on $g^2$. $S^B$ 
also generates the components $\tau_n^{(B)}(k^2)$, 
$\sigma_n^{ij(B)}(k^2)$ , with forms similar to that of 
$\pi_n^{(B)}(k^2)$. The leading 
divergent parts, 
$\pi_{nL}^{(B)}$ etc., are similar to those in (A.4). 

In Case A1, we have, firstly, the set $S^{(1)}$ of $n$-loop diagrams, 
of {\it ffb\/} 
vertices, counterterm insertions, boson mass insertions 
and an even number of fermion mass insertions, such that 
the order is $O(g^{2n})$; which gives, as in Case A2, the component
\be
\pi_{2n}^{(1)} (k^2) = g^{2n} [\pi_{2n,n}^{(1)}\eps^{-n} +\pi_{2n,n-
1}^{(1)} (k^2) \eps^{-n+1} +\cdots ], 
\ee
together with $\tau_{2n}^{(1)}(k^2)$, $\sigma_{2n}^{ij(1)}(k^2)$, of 
similar 
forms, and leading divergent parts $\pi_{2nL}^{(1)}$, 
$\tau_{2nL}^{(1)}$, $\sigma_{2nL}^{ij(1)}$ of the forms of those in 
(A.4). In addition, there is 
the set of $S^{(1+)}$ diagrams that contain {\it ffb\/} vertices, 
counterterm 
vertices, boson mass insertions and 
an odd number of fermion mass vertices, such that the order is 
$O(g^{2n+1})$,\ $n\ge 1$. 
The set $S^{(1+)}$ does not contain diagrams that consist only of {\it
ffb\/}
 vertices, so that corresponding sets $S^0$, $S^{00}$, and $S^0$ 
components 
(A.1), do not exist, with the result that the first or first several 
$O(\eps^{-n}), O(\eps^{-n+1}),\ldots$ terms in the $\pi$, $\tau$, 
$\sigma$ components 
might be zero (the most divergent term is $O(\eps^{-n})$, at most, 
because of the argument preceding 
(A.3), above). Accordingly, we assume that in Case A1, the
 $O(g^{2n+1})$ components of $\pi(k^2)$, $\tau(k^2)$ and 
$\sigma^{ij}(k^2)$ 
are of the form of
\begin{equation} \pi_{2n+1}^{(1)}(k^2) = g^{2n+1} [ 
\pi_{2n+1,n}^{(1)} \eps^{-n} + \pi_{2n+1,n-1}^{(1)} (k^2)\eps^{-
n+1} +\cdots], \end{equation}
where $\pi_{2n+1,n}^{(1)}$ is real and independent of $k^2$, and 
again we define leading divergent parts 
\bq
\pi_{(2n+1)L}^{(1)} &=& g^{2n+1} \pi_{(2n+1),n}^{(1)} 
\eps^{-n},\qquad
\tau_{(2n+1)L}^{(1)} = g^{2n+1} \tau^{(1)}_{(2n+1),n} \eps^{-n}, 
\nonumber \\
\sigma_{(2n+1)L}^{ij(1)} 
&=& g^{2n+1} \sigma^{ij(1)}_{(2n+1),n} \eps^{-n} \ ;
\eq
however, these $O(\eps^{-n})$ terms 
and some of the succeeding 
$O(\eps^{-n+1}),\ldots$ terms in (A.7) and the similar series for 
$\tau_{2n+1}^{(1)}(k^2)$, $\sigma_{2n+1}^{(ij(1)}(k^2)$, might be 
zero.  

We assume that we have forms similar to those above for the self-
energy 
component functions $a_1(p^2),\, b_1(p^2)$ for $\psi$ and 
$a_2(p^2),\, b_2(p^2)$ for $\eta$ (see section~4), e.g.\ for 
$a_1(p^2)$ we have, as developments from the $S^0$-set assumption 
(A.2), 
\be \left.
\begin{array}{l}
a_{1,2n}^{(2)}(p^2) 
= g^{2n} [a_{1,2n,n}^{(2)} \eps^{-n} 
+ a_{1,2n,n-1}^{(2)} (p^2) \eps^{-n+1} +\cdots ],\\ \noas
a_{1n}^{(B)}(p^2) 
= w^{n} [a_{1,n,n}^{(B)} \eps^{-n} 
+ a_{1,n,n-1}^{(B)} (p^2) \eps^{-n+1} +\cdots ],\\  \noas
a_{1,2n}^{(1)}(p^2) 
= g^{2n} [a_{1,2n,n}^{(1)} \eps^{-n} +\cdots ],\\        \noas
a_{1,2n+1}^{(1)} (p^2) 
= g^{2n+1} [a_{1,2n+1,n}^{(1)} \eps^{-n}+\cdots], \end{array} \qquad
\right\}
\ee
with the leading divergent parts
\be 
a_{1,2nL}^{(2)} = g^{2n} a_{1,2n,n}^{(2)} \eps^{-n},\ldots
\ee
real and independent of $p^2$.

For the general vertex part $g\hbar^{-1}\gamma^\mu[ V_1(p,p') 
+\gamma^5 V_5(p,p')]$  
at a $\bar\psi\psi$-boson or $\bar\eta\eta$-boson vertex, we assume 
similarly that each of $V_1$, $V_5$ consists of components of the 
forms (writing 
$V$ for $V_1$, $V_5$), in Cases A1, A2 and B,
\be\left.
\begin{array}{l}
V_{2n}^{(1)}(p,p') 
= g^{2n} [V_{2n,n}^{(1)}\eps^{-n} + 
V_{2n,n-1}^{(1)} (p',p')\eps^{-n+1} +\cdots ],\\ \noas
V_{2n+1}^{(1)}(p,p') 
= g^{2n+1} [V_{2n+1,n}^{(1)}\eps^{-n} + V_{2n+1,n-1}^{(1)} 
(p',p')\eps^{-n+1} 
+\cdots ],\\ \noas
V_{2n}^{(2)}(p,p')
= g^{2n} [V_{2n,n}^{(2)} \eps^{-n} + \cdots],\\ \noas
V_n^{(B)}(p,p')
= w^n [V_{n,n}^{(B)} \eps^{-n} +\cdots ],
\end{array} \quad\right\}\quad
\ee
in which $V_{2n,n}^{(1)}$ etc.\ are real and independent of $(p,p')$. 
The leading divergent parts are defined by (A.19), (A.20), below. 
For $V_{2n+1}^{(1)}(p,p')$, the leading divergent part 
$g^{2n+1} V_{2n+1,n}^{(1)} \eps^{-n}$ and some of the succeeding 
terms 
might be zero.

We use the notations
\be
\pi'(k^2) = \frac{\partial}{\partial k^2} \pi(k^2), 
\quad \pi'' = \frac{\partial} {\partial k^2} \pi',\ldots,
\ee
\be
\hat\pi = \pi(\calm^2), \quad \hat\pi' = \pi'(\calm^2),\ldots,
\ee
and similarly for $\tau$, $\rho$, $a(p^2)$, $b(p^2)$. From the 
$k^2$-independence of 
the leading divergent parts $\pi_{2nL}^{(1)}$, etc., we see that
\bq
\hat\pi_{2nL} ^{(1)} &=& \pi_{2nL}^{(1)}, \qquad 
\hat\pi_{(2n+1)L}^{(1)}= \pi_{(2n+1)L}^{(1)},\nonumber\\
\hat\pi_{2nL} ^{(2)} &=& \pi_{2nL}^{(2)}, \qquad 
\hat\pi_{nL}^{(B)}= \pi_{nL}^{(B)},
\eq
and that similar relations hold for $\tau$ and $\sigma$, and so for 
$\rho$. 

At (27) we introduced the notation $\hat\pi_{qL}^{(J)}$, 
$c_{3q}^{(J)}$, 
$\sigma_{qL}^{(ij)(J)}$, where $J=1,2,B$ labels the case, and $q=2n$ 
for $J=2,B$, while $q=2n,2n+1$ for $J=1$. At (46) we introduced 
$\tilde\pi_{q,n}$. It is also convenient to introduce
\be
\tilde\pi_L^{(J)} = \pi_L^{(J)} -\hbar^{-1}c_3^{(J)} = \sum_n 
(\pi_{qL}^{(J)} -\hbar^{-1}c_{3(q)}^{(J)}), \ee
\be
\hat a_{jL}^{(J)} = a_{jL}^{(J)} = 
\left\{\quad \begin{array}{l}
\sum(g^{2n} a_{j,2nL}^{(1)} +g^{2n+1} a_{j,(2n+1)L}^{(1)}) 
\epsilon^{-n}\\
\sum g^{2n} a_{j,2nL}^{(2)} \epsilon^{-n} \\
\sum w^n a_{j,nL}^{(B)} \epsilon^{-n} \end{array}\quad\right\},\ee
and
\bq
b_{jL}^{(J)} &=& \sum m_i m_k \sigma_{jL}^{(ik)(J)} \nonumber\\
&=& \left\{\quad 
\begin{array}{l}
\sum(g^{2n} b_{j,2nL}^{(1)} +g^{2n+1} b_{j,(2n+1)L}^{(1)}) 
\epsilon^{-n}\\
\sum g^{2n} b_{j,2nL}^{(2)} \epsilon^{-n} \\
\sum w^n b_{j,nL}^{(B)} \epsilon^{-n} \end{array}\quad\right\},\eq
in which 
\be a_{j,2nL}^{(2)} = a_{j,2nL(1)}^{(2)} + 
a_{j,2nL(5)}^{(2)}\gamma^5, \ee
and similarly for $a_{j,2nL}^{(1)}$, $a_{j,(2n+1)L}^{(1)}$ and the 
other $a,b$ quantities. The forms of $\sigma_{jL}^{(ik)(J)}$ are 
evident from (A.17). In a similar way, we write
\be 
V_{jL(1)}^{(J)} = \left\{\quad \begin{array}{l}
\sum(g^{2n} V_{j,2n,n(1)}^{(1)} +g^{2n+1} V_{j,2n+1,n(1)}^{(1)}) 
\epsilon^{-n}\\
\sum g^{2n} V_{j,2n,n(1)}^{(2)} \epsilon^{-n} \\
\sum w^n V_{j,n,n(1)}^{(B)} \epsilon^{-n} \end{array}\quad 
\right\},\ee
and
\be 
V_{jL(5)}^{(J)} = \left\{\quad \begin{array}{l}
\sum(g^{2n} V_{j,2n,n(5)}^{(1)} +g^{2n+1} V_{j,2n+1,n(5)}^{(1)}) 
\epsilon^{-n}\\
\sum g^{2n} V_{j,2n,n(5)}^{(2)} \epsilon^{-n} \\
\sum w^n V_{j,n,n(5)}^{(B)} \epsilon^{-n}. \end{array}\quad 
\right\}.\ee
As discussed in the main text, some of $a_{j,(2n+1)L}^{(1)}$, 
$b_{j,(2n+1)L}^{(1)}$, $V_{j,2n+1,n(1)}^{(1)}$, $ 
V_{j,2n+1,n(5)}^{(1)}$ might be zero.

\bigskip
\renewcommand{\theequation}{B.\arabic{equation}}
\renewcommand{\thesubsection}{B.\arabic{subsection}}
\setcounter{subsection}{0}
\setcounter{equation}{0}
\appendix{\bf Appendix B}

\subsection{Loop by loop renormalisation of the boson propagator}

Following the renormalisation of the boson propagator to one loop in 
Case A2 in (29) to (36), we renormalise up to two loops. We obtain
\bq 
e_2 \!&=& \![1+\hbar \hat\pi_{2L}^{(2)}-c_{3(2)} + \hbar 
\hat\pi_{4L}^{(2)} - c_{3(4)}] M^2 \nonumber\\
&\quad& - \hbar(\hat\rho_{2L}^{(2)}+\hat\rho_{4L}^{(2)}) +O(\eps^{-
1})\qquad \\ 
&=& O(\eps^{-1}) \eq 
on imposing (27), and 
\bq (\Omega_3^{-1})^{}_2 &=& 1 +\hbar \hat\pi_{2L}^{(2)} -
c_{3(2)}^{(2)} 
+\hbar \pi_{4L}^{(2)} -  c_{3(4)}^{(2)} +O(\eps^{-1}) \\
&=& O(\epsilon^{-2}).
\eq 
$R_2(k^2)$ is $O(\eps^{-1})$, so that (34), (35), (36) hold up to 
two loops. 
Proceeding in this way, we see that (36) holds to $N$ loops, $N$ 
arbitrarily large. A parallel treatment of Case B gives the same 
result, that (36) holds to $N$ loops.  

For Case A1 at one loop, the development, now in 
$\hat\pi_2^{(1)}$, $\hat\rho_2^{(1)}$, parallels that of Case A2, 
from (29), to reach (36) again. Then up to (1+)-loops, we have, 
using (A.8), 
\be 
e_{1+} = (1+\hbar \hat\pi_{2L}^{(1)} -c_{3(2)} 
+\hbar \hat\pi_{3L}^{(1)} -c_{3(3)} M^2 -
\hbar(\hat\rho_{2L}^{(1)} +\hat\rho_{3L}^{(1)}) +O(\eps^0),\ee
\be
(\Omega_3^{-1})_{1+} = 1+\hbar \hat\pi_{2L}^{(1)} -c_{3(2)} 
+\hbar \hat\pi_{3L}^{(1)} -c_{3(3)} +O(\eps^0),\ee
\be
R_{(1+)}(k^2) = 
O(\eps^0).  \ee
Imposing (27), we see (using (A.8)) that, as for 
one loop, $e_{1+}$ is $O(\eps^0)$ and $(\Omega_3^{-1})_{1+}$ is 
$O(\eps^{-1})$, whether or not $\hat\pi_{3L}^{(1)}$ is zero. 
Accordingly we obtain (36) again, to (1+)-loops in Case A1. Up to 
two loops, we find that (as in Case A2) $e_2$ is reduced from 
$O(\eps^{-2})$ to $O(\eps^{-1})$ by (27), while $(\Omega_3^{-1})_2$ 
is $O(\eps^{-2})$ and $R_2(k^2)$ is $O(\eps^{-1})$, which leads 
again to (36). Up to (2+)-loops, the argument is similar to that 
for (1+)-loops, with $e$ reduced to $O(\eps^{-1})$ and $\Omega_3^{-
1}$ of $O(\eps^{-2})$, $R(k^2)$ of $O(\eps^{-1})$, to give (36) 
again. We proceed in this way to any number of loops.

\subsection{Order by order renormalisation of the boson propagator}

In (42) to (45) we developed the renormalisation of the boson 
propagator to $O(g^2)$, $O(w)$ in Cases A1, A2 and B. 

In Cases A2 and B we proceed order by order in $g^2,w$. Up to 
$O(g^4)$ in A2, $O(w^2)$ in B, we have 
\begin{equation} 
e_{g4}^{(2)} = B^2 g^4(1-i\delta), \quad e_{w2}^{(B)} = B^2 
w^2(1-i\delta), \ee which are $O(\eps^0)$, and 
\bq (\Omega_3^{-1})_{g4}^{(2)} &=& 1+g^2 \tilde\pi_{2,1}^{(2)} 
\eps^{-1} +g^4 
\tilde\pi_{4,2}^{(2)} \eps^{-2} +O(\eps^{-1},\eps^0),\\ 
(\Omega_3^{-1})_{w2}^{(B)} &=& 1+w \tilde\pi_{1,1}^{(B)} 
\eps^{-1} +w^2 \tilde\pi_{2,2}^{(B)} \eps^{-2} +O(\eps^{-
1},\eps^0), \eq 
(using the notation (46)), which are $O(\eps^{-2})$, while 
$R_{g4}^{(2)}$, 
$R_{w2}^{(B)}$ are $O(\eps^{-1})$. Again we obtain (36) to 
these orders. Similarly, up to $O(g^6)$, $O(w^3)$ we have 
\be 
e_{g6}^{(2)} = (1+g^2\tilde\pi_{2,1}^{(2)}\eps^{-1}) B^2g^4 (1-
i\delta) -\sum (\beta_i \beta_j g^4 ) ( g^2 
\hat\sigma_{2,1}^{ij(2)} \eps^{-1}),\ee
\be
e_{w3}^{(B)} =
(1+w\tilde\pi_{1,1}^{(B)}\eps^{-1}) B^2w^2 (1-i\delta) 
-\sum (\beta_i \beta_j w^2) (w \hat\sigma_{1,1}^{ij(B)} 
\eps^{-1}),\ee
\be
(\Omega_3^{-1})_{g6}^{(2)} = 
1+g^2\tilde\pi_{2,1}^{(2)}\eps^{-1} 
+g^4\tilde\pi_{4,2}^{(2)}\eps^{-
2}+g^6\tilde\pi_{6,3}^{(2)}\eps^{-3} +O(\eps^{-2},\eps^{-
1},\eps^0),\ee
\be 
(\Omega_3^{-1})_{w3}^{(B)} = 
1+w\tilde\pi_{1,1}^{(B)}\eps^{-1} 
+w^2\tilde\pi_{2,2}^{(B)}\eps^{-2}+
w^3\tilde\pi_{3,3}^{(B)}\eps^{-3}+O(\eps^{-2},\eps^{-
1},\eps^0), \ee 
and $R_{g6}^{(2)}$, $R_{w3}^{(B)}$ are 
$O(\eps^{-2})$. Again we obtain (36).  

For Case A1, the treatment up to $O(g^2)$ is given in (42) 
to (45). Up to $O(g^3)$ we have \bq e_{g3}^{(1)} &=& e_{g2}^{(1)} 
= B^2 g^2 (1-i\delta),\\ 
(\Omega_3^{-1})_{g3} &=& 1+g^2 
\tilde\pi_{2,1}^{(1)} \eps^{-1} +g^3 \tilde\pi_{3,1}^{(1)}\eps^{-
1} +O(\eps^0) \eq 
and $R_{g2}^{(1)}$, $R_{g3}^{(1)}$ are 
$O(\eps^0)$, so that we obtain (34), (35) and (36) to this order. 
Note that going to $O(g^{2n+1})$ does not increase the degree of 
divergence  from that at $O(g^{2n})$ (see (A.6), (A.7)). Up to 
$O(g^4)$ 
we have 
\be e_{g4}^{(1)} = (1+g^2\tilde\pi_{2,1}^{(1)}\eps^{-
1}) B^2g^2 (1-i\delta) -\sum (\beta_i \beta_j g^2) ( g^2 
\hat\sigma_{2,1}^{ij(1)} \eps^{-1}) +O(\eps^0),\ee
\be (\Omega_3^{-
1})_{g4}^{(1)} = 1+g^2\tilde\pi_{2,1}^{(1)}\eps^{-1} 
+g^3\tilde\pi_{3,1}^{(1)}\eps^{-
1}+g^4\tilde\pi_{4,2}^{(1)}\eps^{-2} +O(\eps^{-1},\eps^0) \ee
and 
$R^{(1)}_{g4}$ is $O(\eps^{-1})$. We obtain (36) again. We continue 
to $O(g^5)$, and so on.

\bigskip
\renewcommand{\theequation}{C.\arabic{equation}}
\setcounter{equation}{0}
\appendix  {\bf Appendix C. Counterterm parameters for the $\psi\psi 
A$ vertex}
\bigskip 

To find the values of $c_1,c_{1(5)}$ that give the result (91), we 
write (89) as
\be
\frac{\alpha^2}{ \Omega_{3L}^{-1} [(\Omega_{2L}^{-1})_{(1)}]^2 } =1, 
\qquad
\frac{\beta^2} { \Omega_{3L}^{-1} [(\Omega_{2L}^{-1})_{(5)}]^2 } =1,
\ee
with the positive square root understood. In Case A2, we write
\be
\alpha = 1+\sum \alpha_{2n} g^{2n}\epsilon^{-n}, \qquad
\beta = \sum \beta_{2n} g^{2n}\epsilon^{-n}, \ee
\be
\Omega_{3L}^{-1} = 1 + \sum p_{2n} g^{2n} \epsilon^{-n}, \qquad
\Omega_{2L}^{-1} = 1+\sum (r_{2n} + t_{2n}\gamma^5) g^{2n} 
\epsilon^{-n}, 
\ee
where
\be
p_{2n} = \pi^{(2)}_{1,2n,n}\ , \ee
\be
r_{2n}+t_{2n}\gamma^5 = a_{1,2n,n}^{(2)} -s_{1,2n}^{(2)} = 
b_{1,2n,n}^{(2)}\ . \ee
The first equation in (C.1) is then
\be
\frac{1+2\alpha_2 g^2\epsilon^{-1} +(2\alpha_4 +\alpha^2_2) 
g^4\epsilon^{-2} 
+(2\alpha_6 +2\alpha_4\alpha_2) g^6\epsilon^{-3} +\cdots}
{1+(p_2+2r_2) g^2\epsilon^{-1} +(p_4 +2p_2 r_2 +2r_4 +r_2^2) g^4 
\epsilon^{-2}+\cdots} =1. \ee
Since
\be
\left[ \frac{1+\cdots +P g^{2N}\epsilon^{-N}}{1+\cdots+ Q 
g^{2N}\epsilon^{-N}}\right]_{O(g^{2N})} 
= \frac{1+\cdots P g^{2N}\epsilon^{-N}}{1+\cdots Q g^{2N}\epsilon^{-
N}} +O(g^{2N+2}), \ee
where the $O(g^{2N})$ subscript denotes ``up to $O(g^{2N})$'', we can 
satisfy 
(C.6) order by order, correctly to any given order $O(g^{2N})$, by 
making the numerator and denominator identical order by order, i.e.\  
by taking
\be\left.
\begin{array}{l}
\alpha_2 = \frac12 p_2 +r_2, \\
\alpha_4 = \frac12(p_4 +p_2 r_2 +2r_4 -\frac14 
p_2^2),\end{array}\quad \right\} \ee
(where we have used the expression for $\alpha_2$ in writing 
$\alpha_4$) and so on. In a similar way, the second equation in (C.1) 
leads to
\be\left.
\begin{array}{l}
\beta_2 = t_2, \\
\beta_4 = t_4 +\frac12 t_2 p_2, \end{array}\quad
\right\}\ee
and so on. Then from (87), (88), (C.2), (C.8), (C.9) we obtain
\bq
c_1 &=& V_{L(1)} +1-\alpha +f\beta \nonumber\\
&=& V_{L(1)} +(-{\tes\frac12} p_2 -r_2 +f t_2)g^2\epsilon^{-
1}\nonumber\\ 
&\quad& + [ -{\tes\frac12}p_4 -{\tes\frac12} p_2 t_2 -r_4 
+{\tes\frac18} p_2^2 +f(t_4 +{\tes\frac12}p_2t_2) ] g^4\epsilon^{-2} 
+\cdots,\qquad \\
c_{1(5)} &=& f^{-1} V_{L(5)} +1-\alpha +f^{-1}\beta \nonumber\\
&=& f^{-1} V_{L(5)} + (-{\tes\frac12} p_2r_2 +f^{-1} 
t_2)g^2\epsilon^{-1}\nonumber\\
&+& [ -{\tes\frac12} p_4 -{\tes\frac12} p_2t_2 -r_4 +{\tes\frac18} 
p_2^2 +f^{-1} (t_4+{\tes\frac12}p_2t_2)] 
g^4\epsilon^{-2} +\cdots,\quad
\eq
where $V_{L(1)}$ and $V_{L(5)}$ are, for the $\psi\psi A$ vertex in 
Case A2, 
$V_{1L(1)}^{(2)}$ and $V_{1L(5)}^{(2)}$, of the form given by (A.19), 
(A.20). 
These values of $c_1 = c_{1(j=1)}^{(2)}$, $c_{1(5)} = 
c_{1(j=1)(5)}^{(2)}$, 
series in $g^{2n}\epsilon^{-n}$, renormalise the $\psi\psi A$ vertex 
matrix to 
the form (91) in Case A2.

We carry through a similar treatment of the $\psi\psi A$ vertex in 
Case B, with 
the first equation in (C.3) replaced by 
\be
(\Omega_{3L}^{-1})^{(B)} = 1+\sum p_n^{(B)} w^{n+1} \epsilon^{-n} 
\ee
and with similar expansions in $w$ for $(\Omega_{2L}^{-1})^{(B)}$, 
$\alpha^{(B)}$ and $\beta^{(B)}$ (the powers of $g$ are still 
present, in the 
coefficients $p_n^{(B)}$, etc.).  We obtain expressions for 
$c_{1(j=1)}^{(B)}$, $c_{1(j=1)(5)}^{(B)}$ analogous to those for 
$c_{1(j=1)}^{(2)}$, $c_{1(j=1)(5)}^{(2)}$ given by (C.10), (C.11). A 
similar treatment goes through for Case A1, involving powers 
$g^{2n}$,\ $g^{2n+1}$. 

\begin{center} ***
\end{center}


\end{document}